\documentstyle[preprint,aps]{revtex}
\begin{document}
\input epsf
\title{Newtonian Quantum Gravity}
\author{K. R. W. Jones\thanks{Expanded version
of a talk presented at the {\em Inaugural Australian
General Relativity Workshop\/}, CMA, A.N.U., Canberra,
September 30th 1994 (to app. {\em Australian
Journal of Physics}, Nov. '95).}\\
Physics Department, University of Queensland,\\
St Lucia 4072, Brisbane, Australia.}
\maketitle
\begin{abstract} \footnotesize
We develop a nonlinear quantum theory
of Newtonian gravity consistent with an objective
interpretation of the wavefunction. Inspired by
the ideas of Schr\"{o}dinger, and Bell, we seek
a dimensional reduction procedure to map complex
wavefunctions in configuration space onto a family
of observable fields in space--time. Consideration
of quasi--classical conservation laws selects the
reduced one--body quantities as the basis for an
explicit quasi--classical coarse--graining. These
we interpret as describing the objective reality
of the laboratory. Thereafter, we examine what
may stand in the role of the usual Copenhagen
observer to localize this quantity against
macroscopic dispersion. Only a tiny change
is needed, via a generically attractive
self--potential. A nonlinear treatment
of gravitational self--energy is thus
advanced. This term sets a scale for
all wavepackets. The Newtonian
cosmology is thus closed, without need
of an external observer. Finally, the
concept of quantization is re--interpreted
as a nonlinear eigenvalue problem. To
illustrate, we exhibit an elementary
family of gravitationally self--bound
solitary waves. Contrasting this theory
with its canonically quantized analogue,
we find that the given interpretation
is empirically distinguishable, in
principle. This result encourages
deeper study of nonlinear field
theories as a testable alternative
to canonically quantized gravity.
\end{abstract}
\section{Introduction}
The Schr\"{o}dinger interpretation
of the wavefunction (Schr\"{o}dinger 1928, Barut 1988)
would be extremely useful if it were consistent. Then
quantum theory would provide a direct means to address
the problem of particle structure, and an immediate
route to construct generally covariant field theories
founded upon analogy with classical continuum physics.

Of course, there are two long--standing problems
that obstruct consideration of this historical
proposal. Firstly, quantum interactions entangle
states, leading to non--separable densities in
configuration space. Secondly, dispersion is
ever present, so that general localized solutions
are unavailable (Schr\"{o}dinger 1926).
Taken together, these difficulties are
insurmountable in a linear theory, and
the continuum option must fail.

In this paper we show that both difficulties
can be overcome within a larger
theory based upon nonlinear wave--equations.
For this purpose we combine: the mathematics
of Kibble (1978,1979), and Weinberg (1989); the
continuum viewpoint of Schr\"{o}dinger (1928);
the measurement scenario of Penrose (1993);
and the physical mechanism of environmentally induced
decoherence (Zurek 1981,1982,1991).

The argument hinges upon exploiting the verified
classical conservation laws to select an explicit
quasi--classical coarse--graining (cf. Gell-Mann
and Hartle 1993). This approach is closest in
spirit to Bell's idea of {\em beables\/}; some
special class of quantities representing the
reality on which laboratory observations are
founded (Bell 1973). Here reduced one--body
densities and currents are advanced to
play this role.

Nonlinearity is then invoked to obtain a generic
physical mechanism to suppress the macroscopic
dispersion which would otherwise spread out
the observable fields. Thus nonlinearity is
to assume that role now given to observers.
Assuming that an objective theory should
be free of an external observer {\em at
all levels\/}, we are led to constrain
the nonlinearity by demanding this
at the {\em simplest level\/}.

Since gravitation is universally
attractive, and is not directly
tested at the quantum level, we
focus upon it as the means to
suppress dispersion. The outcome is
a possible new avenue to a theory
of individual events --- although,
for simplicity, this paper treats
only Newtonian gravity to secure
the foundation for more general
nonlinear theories.

The key motivation behind this line of enquiry
is the cosmological conundrum posed by quantum
measurement (Bell 1981). Nonlinear theories
offer new modes of physical
interaction that do not quantum entangle.
Since the observer is now treated as a
separable (i.e. non--entangled) participant
in quantum mechanics, nonlinear wave--equations
are an attractive option for the physical
treatment of measurements (Jones 1994b).
However, their interpretation remains
problematic. This suggests that
we look for a specific theory,
using self--consistency and
the question of measurement
as guides (Jones 1995).

For one massive scalar particle with nonlinear
gravitational self--interactions the goal of
observer--free localization is achieved, and
the theory is self--consistent. Correspondence
principle arguments are then used to obtain a
many--particle theory. The gravitational
wave--equation lies within the generalized
dynamics of Weinberg (1989), and resembles
the approximate equations of Hartree--Fock
electromagnetism (Brown 1972).

Quasi--classical coarse--graining plays a central role
in the physical interpretation of these equations according to
the continuum viewpoint of Schr\"{o}dinger (1928). As
soon as the initial obstructions to it are overcome
one realizes that quantum field theory is perhaps
open to a significant reformulation in toto.

For instance, as Barut (1990) has noted in
his attempt to reformulate QED as a linear
theory with nonlinear self--energy terms,
the time--like component of a particle
current need not be of definite sign if
it is a {\em charge density\/}, and not
a {\em probability density\/}. Hence
the Dirac equation might be viewed as
describing a single entity. Then one
might simplify the conceptual basis
of quantum field theories, to free
them from the shackles of
perturbative method, and render
them closed theories
applicable to the entire cosmos. Thus one
may perceive already the outlines for a
program of unification based upon
nonlinearity, the Schr\"{o}dinger
interpretation, and a more physical
approach to quantum measurement.

Since the interpretation of nonlinear theories has
been a major historical stumbling block the early
material, sections $2$ through $6$, is devoted to
addressing this within a continuum philosophy a
l\'{a} Schr\"{o}dinger (1928). Coarse--graining
is dealt with early on, in section $3$. Then, in
section $7$, we generalize Lagrangian dynamics
to complex fields in configuration space.  In
sections $8$ through $12$, we construct the
nonlinear theory of pure Newtonian gravity.
Its relevance is then discussed in the
concluding section.

\section{Continuum physics and nonlinearity}
Schr\"{o}dinger (1928), Einstein (1956),\footnote{Einstein has offered
the following comment on quantized field theories: ``I see in this
method only an attempt to describe relationships of an essentially
nonlinear character by linear methods''} and de Broglie (1960),
were each dissatisfied with the Copenhagen interpretation, and
sought some means to rid quantum mechanics of its reliance upon
the philosophical notion of observers. One possible option
they considered is to introduce nonlinearity, so that the
microscopically tested superpositions are inhibited at
the macroscopic level of the observing instrument. The
difficulty is to locate a candidate nonlinearity, and
to interpret such theories consistently. Here we
reconsider the continuum viewpoint espoused by
Schr\"{o}dinger (1928), and seek to complete
his conception.

The easiest way to appreciate the historical ambiguity
surrounding the interpretation of the wavefunction is
to examine a {\em nonlinear\/} wave--equation, such as
\begin{equation}
\label{nlse}
i\hbar\frac{\partial\psi({\bf x},t)}{\partial t}
= \left\{-\frac{\hbar^{2}}{2m}\nabla^{2}
  \pm\kappa |\psi({\bf x},t)|^{2}\right\}\psi({\bf x},t),
\end{equation}
(later we consider a general self--potential $V[\psi,\psi^{*}]$
in place of the term $\pm\kappa |\psi({\bf x},t)|^{2}$). Such
equations arise frequently as the Hartree--Fock approximation
to an interacting quantum field theory (see e.g., Brown 1972,
Kerman and Koonin 1976). Although of practical utility, they
are generally ignored in discussions of fundamental physics
- - --- primarily because the program of canonical quantization
leads always to a linear theory (Dirac 1958).\footnote{
Recall the logic of this algorithm. Classical Poisson brackets are
replaced by commutators and the correspondence principle becomes
the expression:
$\lim_{\hbar\rightarrow 0} [\hat{f},\hat{g}]/i\hbar = \{f,g\}_{\rm PB}$
(Dirac 1958). Jones (1992, 1994b) has shown that this algebraic
correspondence is incompatible with linearity. The Copenhagen
theory {\em does not\/} contain its classical limit! Thus we
examine the physical correspondence principle afresh and look
for new principles in place of the canonical method.}

Most importantly, the orthodox Copenhagen interpretation demands
linear equations. A measurement might occur at any instant, due
to the intervention of an observer. In virtue of the split,
between dynamics and the observational process, one must
ensure that quantum evolution preserves transition
probabilities. Wigner's theorem asserts that the
only continuous probability preserving maps on
the space of states are linear  unitary (for a
recent review see Jordan (1991)). Linearity for
unobserved systems is thus reconciled with ---
and demanded by --- the postulate of a quantum
jump during measurements.

To go beyond this simplistic idealized picture we need
to recognize the pragmatism which lies at its root. It
is a successful algorithm for conducting computations,
but it ignores the physical basis of measurements. To
probe coherent few--particle dynamics one must first
ensure conditions of isolation. This we describe by
introducing a single wavefunction for the object
system --- neglecting the rest of the Universe.
However, subsequently this isolation must be
broken --- when a measurement is made. At
this stage the theorem of Wigner becomes
suspended, and the collapse postulate
is invoked. In practice, one is then ascending
the scale from the few particles of interest
to the collective behaviour of many within
the measuring apparatus. Therefore, it is
natural to associate any breakdown of the
superposition principle with emergent
many--body nonlinearities.\footnote{The plausibility
of associating nonlinearity with objective theories is made
clearest by Schr\"{o}dinger's example of the cat paradox
(Schr\"{o}dinger 1935). The superposition principle is
clearly incompatible with objective interpretations,
for then Schr\"{o}dinger's cat would be genuinely
alive {\em and\/} dead. The hypothesis of emergent
many--body nonlinearities breaks the logical
chain of extrapolation from micro to macro
physics. Objective nonlinear theories do
not enforce a cat paradox.}

\begin{figure}[p]
\centerline{\epsfbox{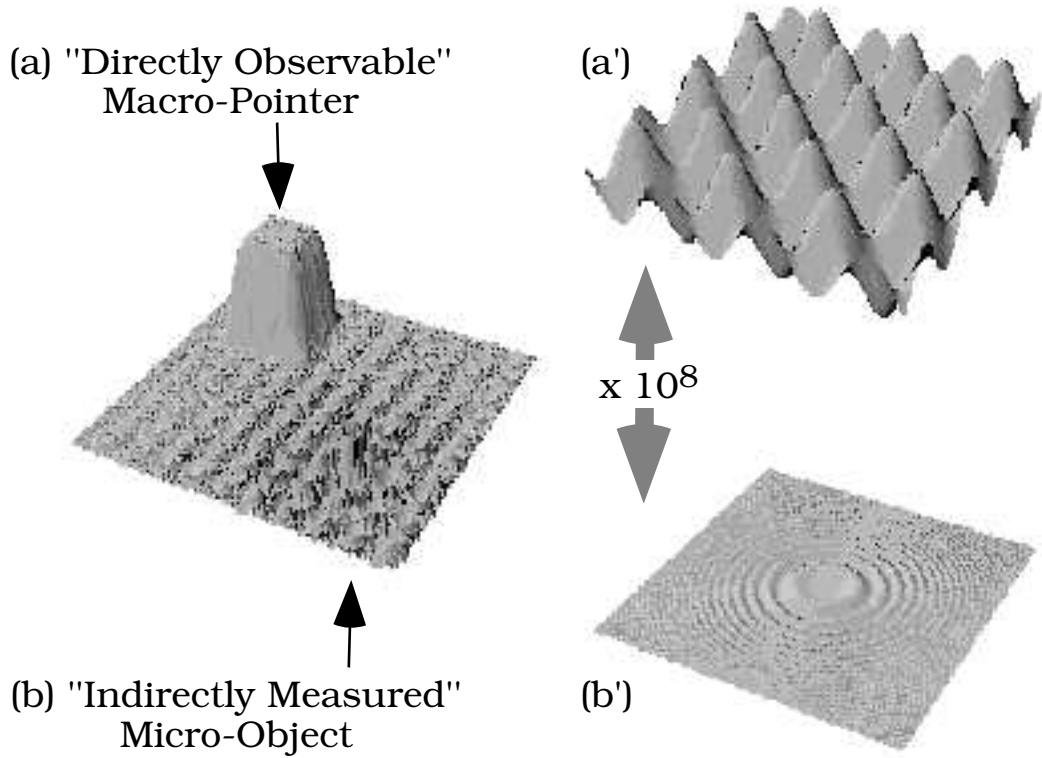}}
\caption{\label{meas} A caricature of the observational process. A directly
observable quasi--classical pointer, item (a), is coupled to a previously
well--isolated elementary constituent, item (b). From the configuration
of (a), and knowledge of the coupling, information can be deduced about
the unseen system (b). In nature, (a) can exhibit gross fluctuations
which are generally interpreted as indicating a discrete foundation
for the quantum world. However, given the vast leap in scale, one
can equally well imagine theories where the observable reality is
described by a coarse--grained continuum field, see the notional
magnifications at (a$'$) and (b$'$). Then the observational process
might be traced to collective correlations and fluctuations in
the values of this field {\em as a whole\/} --- without any
assumption of fundamental discreteness. The key to such a
continuum theory must lie in the chosen prescription for
coarse--graining. It is here that elements of quantum
theory might be brought into direct correspondence
with the observable reality, so to bypass any need
for an observer.}
\end{figure}

Consider the quantum description of a
many--body laboratory instrument, see Fig.~\ref{meas}.
Through the collective evolution of its constituents
it gives concrete form to information gained about
the microscopic world. Although Copenhagen physics
assumes a discrete picture of the microworld, the
quasi--classical level resembles a continuum. Since
our sole experience of the micro--world is founded
in observations at this level (Bohr 1928,1949) one
must admit the logical possibility of a continuum
foundation to all physics. Given the huge gulf
in scale ($\sim 10^{23}$ in number) between
micro and macro physics, a continuum whole
may well fluctuate (when stong fields are
involved) to appear composed of discrete
parts --- the idealized point--like particles
of Copenhagen quantum physics.

Bell (1973) has sought a more objective theory through
his concept of ``beables'', some class of quantities
representing the objective status of observing
instruments. Such beables are readily imagined
as fields defined in ordinary space--time which
describe the totality of the atomic world, and
the observing instrument. The main conceptual
problem is to recover the notion of elementary
particles, as separate well--localized entities.
The continuum field must appear composed of
fluctuating indivisible parts, and we must
ensure that any fundamental level of
determinism is inaccessible to direct
empirical control.

Here we treat particles as an idealized concept
appropriate to the {\em weak--field limit\/}
when observable continuum fields decompose
into several distinct lumps. Then we may
subsume Copenhagen physics as an ideal
case, and employ it as a known limit
to guide the formulation of new
equations.

Some immediate guidance to a viable coarse--grained
continuum interpretation may be found in practical
many--body equations like (\ref{nlse}). There the
quantity $|\psi|^{2}$ is interpreted as denoting
the density of particles in an aggregate. Examining
(\ref{nlse}), we construct the current
\begin{equation}
{\bf j}({\bf x},t) = \frac{\hbar}{2mi}
\left\{\psi^{*}({\bf x},t)\nabla \psi({\bf x},t)
- - -      \psi({\bf x},t)\nabla \psi^{*}({\bf x},t)\right\}
\end{equation}
and so readily verify the equation of continuity
\begin{equation}
\frac{\partial \rho}{\partial t} + \nabla\cdot {\bf j}({\bf x},t) = 0,
\end{equation}
where $\rho({\bf x},t) =  |\psi({\bf x},t)|^{2}$ (provided
the nonlinear self--potential $V$ is real--valued).

Thus we can interpret {\em macroscopic\/} conserved
currents in two different ways. They could describe
the evolution, in probability density, of very many
point--like particles spread as $\rho({\bf x},t)$,
or the flow of continuum matter.
Recall, both Born and Schr\"{o}dinger, each
employed current conservation to defend their
respective interpretations. With the ancient
Greeks, we may ask again: {\em Does nature
give preference to the discrete or
continuum viewpoint?\/}.

\section{Explicit quasi--classical coarse graining}
So what are the problems with the continuum viewpoint?
Firstly, the candidate for a macroscopic ``beable'' continuum
field must be generated from a wave in configuration space.
Secondly, we must explain how a physics based upon this
assumption could exhibit the discrete and stochastic
behaviour we observe in quantum experiments.

Evidently, the key avenue into such a theory must lie within
that role now played by the Copenhagen observer. As soon as
we drop the Born interpretation something must be found in
place of the observer. Presently, this agent serves to
connect wavefunctions with the laboratory reality.
Therefore, the continuum coarse--graining must effect
a similar mapping from the abstract mathematics into
the concrete world of observations.

Since wavefunctions are generally formulated as fields on
a configurational space--time, the mapping into observable
fields must reduce fields in $N(3+1)$ dimensions down to
the familiar $3+1$ dimensions apparent in the laboratory.
Hence we conceive of an observable level whose dynamics
is determined at an unobservable level, with the two
levels linked by dimensional reduction. The idea is
perhaps familiar already from string theory, where
embarassing dimensions are compactified away, and
one has the sophistication to recognize that
mere mathematics is not reality, but can
serve our efforts to model it.

Many such prescriptions are possible, so we must adopt
some principle to guide the search. The avenue we take
is to look for quasi--classical conservation laws, and
confine attention to mappings which recover these. The
rationale is clear: {\em observable fields must
display the observed conservation laws\/}.

Consider, therefore, the many--body Schr\"{o}dinger
equation for $N$ particles, each of mass $m_{j}$.
In accordance with established physics we describe
these by a wavefunction $\Psi({\bf X};t)$, where
${\bf X} = ({\bf x}_{1},\ldots,{\bf x}_{N};t)$
is a $3N$--vector in configuration space. The
corresponding Schr\"{o}dinger equation reads
\begin{equation}
\label{generic}
i\hbar\frac{\partial\Psi({\bf X};t)}{\partial t}
= - \sum_{j=1}^{N}
\frac{\hbar^{2}}{2m_{j}}\nabla^{2}_{{\bf x}_{j}}\Psi({\bf X};t)
+ V[\Psi,\Psi^{*}]\Psi({\bf X};t),
\end{equation}
where $V[\psi,\psi^{*}]$ is a general real--valued potential.

Higher dimensional conservation laws in configuration space
are easily obtained from (\ref{generic}). For instance, one
may form a $3N$--vector many--body current
\begin{equation}
{\bf j}^{(N)}({\bf X};t)  \equiv
({\bf j}^{(1)}_{1}({\bf X};t),\ldots,{\bf j}^{(1)}_{N}({\bf X};t)),
\end{equation}
composed of $N$ different $3$--vector partial currents
\begin{equation}
\label{partialcurrent}
{\bf j}^{(1)}_{j}({\bf X};t) \equiv
\frac{\hbar}{2m_{j}i}
\left\{\Psi^{*}({\bf X};t)
\nabla_{{\bf x}_{j}}\Psi({\bf X};t) - \Psi({\bf X};t)
\nabla_{{\bf x}_{j}}\Psi^{*}({\bf X};t)\right\},
\end{equation}
along with the corresponding scalar density
\begin{equation}
\label{partialdensity}
\rho^{(N)}({\bf X};t)
\equiv \Psi^{*}({\bf X};t)\Psi({\bf X};t).
\end{equation}
Thus we obtain a many--body equation of continuity
\begin{equation}
\label{manybody}
\partial_{t}\rho^{(N)}({\bf X};t)
+\partial_{\bf X}\bullet
{\bf j}^{(N)}({\bf X};t) = 0,
\end{equation}
where ``$\bullet$'' denotes the natural $3N$--vector dot product.

Such conservation laws would normally be introduced to
discuss the joint measurement of $N$ particles. To do
this in practice would require $N$ detectors, each of
them composed of many particles. Therefore, we put a
particle picture aside, and concentrate upon how one
is to describe the instrumental readout of a notional
meter. The ``needle'' position is encoded jointly in
the collective state of very many particles, and so
a coarse--grained treatment will suffice, see again
Fig.~\ref{meas}. In
classical point mechanics the mass density
\begin{equation}
\rho({\bf x};t) =
\sum_{j=1}^{N} m_{j}\delta({\bf x} - {\bf x}_{j}),
\end{equation}
is one candidate to describe the substance of this
pointer. In classical continuum physics we simply
smear out each of the $N$ delta functions with a
density $\rho_{j}(x_{j})$, and integrate over
configuration space. In analogy with this
familiar case the quantum entanglement
may be submerged from direct view by
integration of the many--body
density (\ref{partialdensity}).

Hence we obtain a {\em reduced one--body density\/}
\begin{equation}
\label{onebodydensity}
{\bf \rho}^{(1)}({\bf x};t) \equiv \int d^{3N}{\bf X}\,
\sum_{j=1}^{N}
\delta ({\bf x} - {\bf x}_{j})
{\bf \rho}^{(N)}({\bf X};t),
\end{equation}
and the corresponding {\em reduced one--body current\/}
\begin{equation}
\label{onebodycurrent}
{\bf j}^{(1)}({\bf x};t) \equiv \int d^{3N}{\bf X}\,
\sum_{j=1}^{N} \delta ({\bf x} - {\bf x}_{j}) {\bf j}^{(N)}({\bf X};t).
\end{equation}
In Copenhagen physics one would recognize these as
describing the probability density and current for
``finding'' all $N$ particles at once, neglecting
the correlations among them. This is exactly the
kind of coarse--graining we require to describe
quasi--classical instrumental readouts. Further,
by integration of (\ref{manybody}) we obtain
\begin{equation}
\label{onebody}
\partial_{t} \rho^{(1)}({\bf x};t)
+ \partial_{\bf x}\cdot {\bf j}^{(1)}({\bf x};t) = 0,
\end{equation}
a {\em reduced one--body equation of continuity\/}. Hence we
recover the familiar quasi--classical law of (local) mass
conservation, without appeal to an observer.

A suitable general prescription, consistent with these
principles, is the familiar one--body field operator
of standard many--body physics. Specifically, we
consider the quantities
\begin{equation}
\label{onebodyfield}
f({\bf x},t) \equiv \langle\Psi|\hat{F}({\bf x},t)|\Psi\rangle
\end{equation}
where $|\Psi\rangle$ is the many--body state, and a
typical one--body operator may be written
\begin{equation}
\hat{F}({\bf x},t) =
\hat{\Psi}^{\dagger}({\bf x},t)F({\bf x},t)\hat{\Psi}({\bf x},t),
\end{equation}
with $\hat{\Psi}^{\dagger}({\bf x},t)$, and $\hat{\Psi}({\bf x},t)$ the
usual field creation and annihilation operators and
$F({\bf x},t)$ is generally a $c$--number
distribution (e.g. $-i\hbar\nabla$ for momentum).
Thus we arrive at an explicit prescription for
what Gell--Mann and Hartle (1993) have referred
to as a quasi--classical coarse graining of the
quantum micro--reality.

\section{Hypothesis of restricted observables}
To complete the interpretation we advance a simple
hypothesis: {\em Only reduced one--body quantities
are directly observable --- all other observations
must be derived from these.\/}\footnote{The clearest antecedent
I am familiar with is Bell's notion of {\em beables\/};
hence his line ``Observables are made out of beables''
(Bell 1973). There is also an obvious similarity
with the de Broglie--Bohm theory of {\em pilot waves\/}
(for a review, and commentary, see Bell 1976), except
that we introduce no extraneous variables to track
point--like particles within the wave. We choose
this hypothesis as a simple means to address the
dilemma of configuration space. The wave--particle
duality of Bohr (1928) is then broken in favour
of waves, with the particle--like properties
treated as an idealization.}
The purpose behind this postulate is to secure the
foundations for a continuum theory, in which there
is no need to introduce an external observer that
must ``find'' the many particles which comprise
a macroscopic instrumental readout.

Now the correspondence principle of Bohr (1928) is
met directly in the construction of the observable
level along with the non--classical property of
quantum entanglement. Via coarse--graining this
property becomes reconciled with our direct
experience of apparently continuous quantities
formulated in a $3+1$ dimensional space--time. However,
unlike classical continuum physics, the dynamics of
the observable fields drawn above is not determined
by {\em their\/} values, but rather those of $\Psi$.

Obviously, quantum non--separability remains important
in determining the behaviour of such reduced fields,
but we are free of the need to explain it within
the reductionist model of particles. The quantum
world is restored to a whole, although it can
resemble a collection of parts whenever the
quantum correlations among these are negligible.

For instance, under isolated conditions, e.g. a diffuse gas,
the many--body wavefunction can be well--approximated by a
factored product (Brown 1972)
\begin{equation}
\Psi({\bf X};t)
\approx \prod_{j=1}^{N} \psi_{j}({\bf x}_{j}).
\end{equation}
The one--body density reduces to a simple classical
sum of $N$ terms $\rho_{j}({\bf x}) = |\psi_{j}({\bf x})|^{2}$,
and these behave as independent extended ``particles''. Unlike in classical
physics, the indistinguishability of identical ``particles''
now follows in consequence of the fundamental hypothesis.
Taking a symmetrized wavefunction
\begin{equation}
\Psi({\bf X};t)
= \frac{1}{\sqrt{N!}} \sum_{\cal P}(\pm)^{\cal P}
\Psi\left({\cal P}[{\bf x}_{1},\ldots,{\bf x}_{N}];t\right),
\end{equation}
with ${\cal P}$ denoting permutations, and sign
$+$ for bosons, and $-$ for fermions, one sees
that the one--body density has, in general, $N$
distinct lumps. In Copenhagen physics different
species of particle are distinguished by their
quantum numbers. Exact conservation laws yield
exact quantum numbers (Itzykson and Zuber 1985).
Similar consideration can be applied to fields.
If each coordinate ${\bf x}_{j}$ has identical
one--body conservation laws, then observation
of their sum via (\ref{onebodydensity}) will
offer no means to associate any one of $N$
lumps with a particular coordinate. The
identity of elementary field coordinates
is lost by integration for observable
fields whose form is invariant to
permutations among these.

\section{Unobservable determinism and locality}
Theories of this kind afford a very natural
explanation for the unrepeatable nature of
quantum observations. There are many $\Psi$
which generate similar one--body fields but
these need not evolve identically, since it
is the wavefunction which serves as the
initial condition. Since only reduced
quantities are observable (by hypothesis),
the experimental conditions are inexactly
reproducible, of necessity. The theory
remains deterministic, but only some
approximate causality is testable.\footnote{One is reminded
here of a remark by Heisenberg (1949): ``The chain of cause
and effect could be quantitatively verified only if the whole
universe were considered as a single system --- but then
physics has vanished, and only a mathematical scheme
remains.'' That is true of Copenhagen physics; but
in this approach the mathematical scheme is filled
with additional content. The dimensional reduction
to observable fields enables us to postulate a
psycho--physical parallelism between these and
the objective reality of natural processes.
Unlike classical psycho--physical parallelism,
fundamental limits to observation and control
are imposed by the inaccessability of quantum
initial conditions. Hence the hypothesis of
restricted observables is central to the
internal consistency of objective theories.}

This is an encouraging sign, but it remains to formulate
a clear locality criterion for the dynamics of fields
in configuration space. Evidently, the observational
requirement upon locality is that effective superluminal
communication be ruled out. This demand applies at the
level of the observable fields, but the usual quantum
non--locality persists in non--classical correlations
among observable field values at different space--time
points. The position is similar to the non--local but
non--communicating theory of Bohm and Bub (1966). We
must remember that the Bell inequality exclusions
(Bell 1988) apply only to local hidden variables
theories. How any physical non--locality is
judged depends greatly upon our assumption
of what is observable.

\section{Scenario for the observational process}
Unlike in Copenhagen physics, the reduced one--body
continuum fields are to describe the objective and
observable state of an entire cosmos, including
quantum transitions. A partial explanation for
these is available via the physical mechanism
of decoherence (Zurek 1981,1982,1991). Recall
that the transition probabilities, and the
basis in which a superposition is finally
resolved, are explained therein by appeal
to environmental effects. These wash out
off--diagonal coherences in the density
matrix leaving only diagonal entries in
the so--called Zurek ``pointer basis''
(Zurek 1981). Each of these appears
weighted by the appropriate quantum
transition probability, see Fig.~\ref{scenario}(a).

However, it remains to locate some physical mechanism
that can reduce the many diagonal entries to just one
manifest outcome. Here we take up a suggestion due to
Penrose (1993), that a gravitational nonlinearity may
resolve gross macroscopic superpositions. The idea is
that, while gravitation is very weak, it is sensitive
to the collective state of a very large assembly of
particles. For example, a dead cat and a live cat
generate very different gravitational potentials.
Since gravitation is a localizing force, it may
intervene to choose just one lump before the
situation becomes absurd.

Interestingly, non--entangling nonlinearities can
cross--couple the diagonal entries in a decohered
density matrix --- {\em without\/} demanding any
further tracing over the environment. Therefore,
the two ideas seem strongest when combined. We
suggest a two--step scenario: probabilities
fixed by decoherence, and individual events
selected by a nonlinear instability, see
Fig.~\ref{scenario}(b). This scenario resembles
the gravitational stochastic reduction idea of
Di\'{o}si (1989), except that no assumption
of intrinsic quantum jumps is needed.

\section{Mathematics of nonlinear theories}
Consistent with the preceding interpretation we consider
replacing many--body operator quantized fields by fields
in configuration space. Thus we study generalizations
of the familiar linear dynamics for complex--valued
fields in configuration space.

One is in search of a formalism which enables the
extraction of linear operators in the weak--field
limit of small nonlinearity. This demand reflects
the physical necessity that the structures of the
present theory be recovered in an orderly manner.
That is, just as the metric field introduced in
general relativity subsumes the flat Minkowski
space metric, we must ensure that a nonlinear
formalism is a {\em natural\/} generalization
of the familiar physical structure of linear
operators acting upon a Hilbert space. In
this way both the mathematics and physical
concepts of a nonlinear theory may contract
upon those of its predecessor.

Some guidance is provided by the previous geometrization
of quantum dynamics due to Kibble (1979), along with the
introduction, by Weinberg (1989), of a restriction upon
the allowed Hamiltonians which enforces the requisite
operatorial structure. Here we present a new argument
to constrain a Lagrangian system of dynamics, so that
relativistic extension is possible (Itzykson and Zuber 1985).

Consider a wavefunction in configuration space
with $N$ particle coordinates, and one time
coordinate. We are thus pre--occupied, at
first, with a non--relativistic dynamics.
We seek a prescription for such which may
effectively subsume and generalize the
standard one. To begin we decompose
the complex field into a pair of
real--valued fields
\begin{equation}
\label{decomposition}
\Psi({\bf x}_{1},\ldots,{\bf x}_{N};t)
=\frac{1}{\sqrt{2}}\left(
   \Phi_{\rm R}({\bf x}_{1},\ldots,{\bf x}_{N};t)
+ i\Phi_{\rm I}({\bf x}_{1},\ldots,{\bf x}_{N};t)\right),
\end{equation}
being its real and imaginary parts. As such,
we may employ understanding of the classical
theory of Lagrangian dynamics to obtain a
nonlinear dynamics for complex fields.

Recognize, however, that the classical parallel is
in no way reflected in physical concepts.
The ``classical fields'' drawn above support no
physical dimension, as they are the real and
imaginary parts of a complex quantity. They
are also fields in configuration space, the
like of which was not encountered prior to
the discovery of Schr\"{o}dinger dynamics.

To deal first, in isolation, with the unfamiliar aspect of
configuration space consider the case of a
single real--valued field,
$\Phi({\bf x}_{1},\ldots,{\bf x}_{N};t)$,
and make the obvious identifications:
$\partial_{t} = \partial/\partial t$;
${\bf X} = ({\bf x}_{1},\ldots,{\bf x}_{N})$;
and $\partial_{{\bf X}} =
(\nabla_{{\bf x}_{1}},\ldots,\nabla_{{\bf x}_{N}})$.
A Lagrangian dynamics in configuration space is then
obtained via the classical principle of least action
\begin{eqnarray}
\label{realaction}
\delta I[\Phi]  & \equiv & \delta
\int_{t_{1}}^{t_{2}} dt\,L[\Phi,\delta_{t}\Phi] = 0 \\
\label{realdensity}
L[\Phi,\delta_{t}\Phi] & \equiv &
\int d^{3N}{\bf X}\,
{\cal L}(\Phi,\partial_{t}\Phi,\partial_{{\bf X}}\Phi).
\end{eqnarray}
Taking the variational derivative, we compute
\begin{equation}
\delta I[\Phi] =
\int_{t_{1}}^{t_{2}} dt\,\int d^{3N}{\bf X}
\left\{
\frac{\partial {\cal L}}{\partial \Phi}\delta \Phi({\bf X};t)
+
\frac{\partial {\cal L}}{\partial \left[\partial_{t} \Phi\right]}
 \delta \partial_{t}\Phi({\bf X};t)
+
\frac{\partial {\cal L}}{\partial \left[\partial_{{\bf X}} \Phi\right]}
\cdot\delta \partial_{{\bf X}}\Phi({\bf X};t)\right\}.
\label{realvariation}
\end{equation}
Adopting now the usual endpoint restrictions $\delta \Phi(t_{1}) = 0$,
and $\delta \Phi(t_{2}) = 0$, with $\Phi({\bf X};t)$ vanishing at
spatial infinity for all $t$, we integrate by parts, and transfer
partials, to obtain the required Euler--Lagrange equations for
real--valued fields in configuration space
\begin{equation}
\label{realeulerlagrange}
\frac{\partial {\cal L}}{\partial \Phi}
- - - \partial_{t}\left(
\frac{\partial {\cal L}}{\partial \left[\partial_{t} \Phi\right]}\right)
- - - \partial_{\bf X}\cdot\left(
\frac{\partial {\cal L}}{\partial \left[\partial_{{\bf X}} \Phi\right]}\right)
= 0.
\end{equation}
Further, upon defining the canonically conjugate momentum
\begin{equation}
\label{realmomentum}
\Pi({\bf X};t) \equiv
\frac{\delta}{\delta \left[\partial_{t} \Phi\right]}
L[\Phi,\partial_{t}\Phi] =
\frac{\partial {\cal L}}{\partial \left[\partial_{t} \Phi\right]}
\end{equation}
and making a Legendre transformation to introduce the Hamiltonian functional
\begin{equation}
\label{realhamiltonian}
H[\Phi,\Pi] = \int d^{3N}{\bf X}\, \Pi({\bf X};t)\partial_{t}\Phi({\bf X};t) -
L[\Phi,\partial_{t}\Phi],
\end{equation}
we obtain the real--valued Hamiltonian system
\begin{equation}
\partial_{t}\Phi({\bf X};t) =
+\frac{\delta H[\Phi,\Pi]}{\delta \Pi({\bf X};t)}
\;\;\mbox{and}\;\;
\partial_{t}\Pi({\bf X};t)  =
- - -\frac{\delta H[\Phi,\Pi]}{\delta \Phi({\bf X};t)},
\end{equation}
with the associated Poisson bracket
\begin{equation}
\left\{G[\Phi,\Pi],H[\Phi,\Pi]\right\}_{\rm PB}
\equiv \int d^{3N}{\bf X}\left\{
\frac{\delta G[\Phi,\Pi]}{\delta \Phi({\bf X};t)}
\frac{\delta H[\Phi,\Pi]}{\delta \Pi({\bf X};t)}
- - -
\frac{\delta H[\Phi,\Pi]}{\delta \Phi({\bf X};t)}
\frac{\delta G[\Phi,\Pi]}{\delta \Pi({\bf X};t)}\right\},
\end{equation}
analogous to the standard one (cf. Itzykson and Zuber 1985).

This formalism embraces all manner of conservative nonlinear
wave--equations. However, with the notable exception of the
real--valued gauge fields of quantum field theory, it is
too general a framework for quantum dynamics. Examining
the complex action principle
\begin{equation}
\label{complexaction}
\delta I[\Psi] = \delta
\int_{t_{1}}^{t_{2}} dt\,\int d^{3N}{\bf X}\,
{\cal L}(\Psi,\partial_{t}\Psi,\partial_{{\bf X}}\Psi) = 0.
\end{equation}
we acknowlege a novel restriction of purely mathematical
origin. Upon promoting real fields to complex fields we
have implicitly chosen a Lagrangian formulated upon two
{\em coupled\/} real--valued fields $\Phi_{\rm R}$ and
$\Phi_{\rm I}$. Together these fix  both $\Psi({\bf X};t)$  and
its complex conjugate $\Psi^{*}({\bf X};t)$. However,
the general action for doubled fields, namely
\begin{equation}
\label{doubledaction}
\delta I[\Phi_{\rm R},\Phi_{\rm I}]
 = \delta
\int_{t_{1}}^{t_{2}} dt\,\int d^{3N}{\bf X}\,
{\cal L}(\Phi_{\rm R},
\partial_{t}\Phi_{\rm R},
\partial_{{\bf X}}\Phi_{\rm R};
\Phi_{\rm I},
\partial_{t}\Phi_{\rm I},
\partial_{{\bf X}}\Phi_{\rm I}) = 0,
\end{equation}
can violate this property (for a simple finite--dimensional
example, see Jones (1994a)).

One may perceive, in this simple observation, the magnificent
opportunity to fix, once and for all, the precise form of the
mathematical formalism into which all physical content must
be poured. An inclusive nonlinear quantum theory must employ
a generalized dynamics compatible with complex--valued fields.
Evidently, this must be a restriction founded
within complex geometry. Elsewhere, we employed  analyticity
conditions to investigate this question (Jones 1994a). This
characterizes linear quantum dynamics as the {\em analytic\/}
restriction of complex nonlinear dynamics. To go beyond
that demands geometrical arguments which are not tied
to the assumption of analyticity.

Here we present a new argument which affords a complete
characterization of complex nonlinear dynamics. After
a global phase change $\Psi\mapsto e^{i\theta}\Psi$
in (\ref{decomposition}), the real and imaginary
parts of the complex field transform to
\begin{eqnarray}
\tilde{\Phi}_{\rm R}(\theta)
& = & \cos \theta \Phi_{\rm R} - \sin \theta \Phi_{\rm I}\\
\tilde{\Phi}_{\rm I}(\theta)
& = & \sin \theta \Phi_{\rm R} + \cos \theta \Phi_{\rm I}.
\end{eqnarray}
Obviously such a gauge freedom may always be implemented at
the level of the  solutions to a complex dynamical
system (because the two fields are not independent).

Therefore, demand that the  action functional respect
the continuous symmetry
\begin{equation}
\label{sym}
I[\Phi_{\rm R},\Phi_{\rm I}] =
I[\tilde{\Phi}_{\rm R}(\theta),\tilde{\Phi}_{\rm I}(\theta)].
\end{equation}
Differentiating both sides we discover that
\begin{equation}
\label{condition}
0 = \int dt\, \int d^{3N}{\bf X}\,\left\{
\frac{\delta I}{\delta \tilde{\Phi}_{\rm R}({\bf X};t)}
\frac{d \tilde{\Phi}_{\rm R}}{d\theta}
+
\frac{\delta I}{\delta \tilde{\Phi}_{\rm I}({\bf X};t)}
\frac{d \tilde{\Phi}_{\rm I}}{d\theta}\right\}.
\end{equation}
Using $\tilde{\Phi}_{\rm R}'(\theta) = - \tilde{\Phi}_{\rm I}(\theta)$,
and $\tilde{\Phi}_{\rm I}'(\theta) = + \tilde{\Phi}_{\rm R}(\theta)$, and
substituting
\begin{eqnarray}
\frac{\delta }{\delta \Phi_{\rm R}({\bf X};t)}
& = & \frac{1}{\sqrt{2}}
\left(
\frac{\delta }{\delta \Psi({\bf X};t)} + \frac{\delta }{\delta \Psi^{*}({\bf
X};t)}\right)\\
\frac{\delta }{\delta \Phi_{\rm I}({\bf X};t)}
& = & \frac{i}{\sqrt{2}}
\left(
\frac{\delta }{\delta \Psi({\bf X};t)} - \frac{\delta }{\delta \Psi^{*}({\bf
X};t)}\right),
\end{eqnarray}
we obtain the {\em complex compatability conditions\/}
\begin{equation}
\label{compatibility}
\int dt\, \int d^{3N}{\bf X}\,
\frac{\delta I[\Psi,\Psi^{*}]}{\delta \Psi({\bf X};t) }
\Psi({\bf X};t) =
\int dt\, \int d^{3N}{\bf X}\,
\frac{\delta I[\Psi,\Psi^{*}]}{\delta \Psi^{*}({\bf X};t) }
\Psi^{*}({\bf X};t),
\end{equation}
of which the
{\em Kibble--Weinberg homogeneity constraint\/} (Kibble 1978, Weinberg 1989)
\begin{equation}
\label{homogeneity}
\int dt\, \int d^{3N}{\bf X}\,
\frac{\delta I[\Psi,\Psi^{*}]}{\delta \Psi({\bf X};t) }
\Psi({\bf X};t) = I[\Psi,\Psi^{*}] =
\int dt\, \int d^{3N}{\bf X}\,
\frac{\delta I[\Psi,\Psi^{*}]}{\delta \Psi^{*}({\bf X};t) }
\Psi^{*}({\bf X};t),
\end{equation}
is a special case (we will comment upon its utility later).

Provided that the action respects (\ref{compatibility}) the
invariance under a global phase change is guranteed. This
we take to be an intrinsic characterization of complex
dynamical systems, defined via the complex
Euler--Lagrange equations
\begin{equation}
\label{complexeulerlagrange}
\frac{\partial {\cal L}}{\partial \Psi}
- - - \partial_{t}\left(
\frac{\partial {\cal L}}{\partial \left[\partial_{t} \Psi\right]}\right)
- - - \partial_{\bf X}\cdot\left(
\frac{\partial {\cal L}}{\partial \left[\partial_{{\bf X}} \Psi\right]}\right)
= 0\;\;\mbox{\rm and c.c.},
\end{equation}
appearing as a conjugate pair.
The corresponding complex momenta, for there are
now two them, are defined as:
\begin{eqnarray}
\label{complexmomone}
\Pi({\bf X};t) \equiv
\frac{\delta}{\delta \left[\partial_{t} \Psi\right]}
L[\Psi,\partial_{t}\Psi] =
\frac{\partial {\cal L}}{\partial \left[\partial_{t} \Psi\right]},\\
\label{complexmomtwo}
\tilde{\Pi}({\bf X};t) \equiv
\frac{\delta}{\delta \left[\partial_{t} \Psi^{*}\right]}
L[\Psi^{*},\partial_{t}\Psi^{*}] =
\frac{\partial {\cal L}}{\partial \left[\partial_{t} \Psi^{*}\right]}.
\end{eqnarray}
The complex Hamiltonian functionals are then:
\begin{eqnarray}
\label{complexhamone}
H[\Psi,\Pi] \equiv
\int dt\,\int d^{3N}{\bf X}\,
\Pi({\bf X};t)\partial_{t} \Psi({\bf X};t)
- - -L[\Psi,\partial_{t}\Psi],\\
\label{complexhamtwo}
H[\Psi^{*},\tilde{\Pi}] \equiv
\int dt\,\int d^{3N}{\bf X}\,
\tilde{\Pi}({\bf X};t)\partial_{t} \Psi({\bf X};t)
- - -L[\Psi^{*},\partial_{t}\Psi^{*}].
\end{eqnarray}
Hence we arrive at the corresponding Hamiltonian equations:
\begin{eqnarray}
\label{hamA}
\partial_{t} \Psi({\bf X};t) =  +
\frac{\delta H[\Psi,\Pi]}{\delta \Pi({\bf X};t)},
\;\;&\mbox{and}&\;\;
\partial_{t} \Pi({\bf X};t)  =  -
\frac{\delta H[\Psi,\Pi]}{\delta \Psi({\bf X};t)};\\
\label{hamB}
\partial_{t} \Psi^{*}({\bf X};t)  =  +
\frac{\delta H[\Psi^{*},\tilde{\Pi}]}{\delta \Pi({\bf X};t)},
\;\;&\mbox{and}&\;\;
\partial_{t} \tilde{\Pi}({\bf X};t)  =  -
\frac{\delta H[\Psi^{*},\tilde{\Pi}]}{\delta \Psi({\bf X};t)}.
\end{eqnarray}
Generally these four equations may be reduced to two,
say (\ref{hamA}), with (\ref{hamB}) obtained as the
conjugate provided that we elect to define
$L[\Psi^{*},\partial_{t}\Psi^{*}]$
as $L[\Psi,\partial_{t}\Psi]$ subject to the mapping
$\Psi\mapsto\Psi^{*}$. If this is not done, then
it may happen that $\Pi$ and ${\tilde \Pi}$ are
not complex conjugates of one another,
in which case one would select just one pair
of equations anyway.

Returning now to (\ref{compatibility}), we see
that this defines a functional equation for the
Lagrangian. To absorb its content we consider
the typical non--relativistic example
\begin{equation}
H[\Psi,\Psi^{*}]
 = i\hbar \int d^{3N}{\bf X}
 \Psi^{*}({\bf X};t)\partial_{t} \Psi({\bf X};t).
- - - L[\Psi,\partial_{t}\Psi]
\end{equation}
Then $\Pi({\bf X};t) = i\hbar \Psi^{*}({\bf X};t)$, and
the equations (\ref{hamA}) become
\begin{equation}
\label{Wham}
i\hbar\partial_{t} \Psi({\bf X};t)  =  +
\frac{\delta H[\Psi,\Psi^{*}]}{\delta \Psi^{*}({\bf X};t)},
\;\;\mbox{and}\;\;
i\hbar\partial_{t} \Psi^{*}({\bf X};t) = -
\frac{\delta H[\Psi,\Psi^{*}]}{\delta \Psi({\bf X};t)},
\end{equation}
which are those proposed previously by Weinberg (1989).

The condition (\ref{compatibility}) is immediately
satisfied by the time--dependent part. A sufficient
condition for the Hamiltonian functional is then
\begin{equation}
\label{compatibilitytwo}
\int d^{3N}{\bf X}\,
\frac{\delta H[\Psi,\Psi^{*}]}{\delta \Psi({\bf X};t) }
\Psi({\bf X};t) =
\int d^{3N}{\bf X}\,
\frac{\delta H[\Psi,\Psi^{*}]}{\delta \Psi^{*}({\bf X};t) }
\Psi^{*}({\bf X};t).
\end{equation}
As an example, return to (\ref{nlse}) and observe
that this equation may be derived from
\begin{eqnarray}
\lefteqn{H[\Psi,\Psi^{*}] \equiv
\int d^{3}{\bf x}\,
\frac{\hbar^{2}}{2m}
\nabla_{\bf x}\Psi^{*}({\bf x};t)\cdot\nabla_{\bf x}\Psi({\bf x};t) }
\nonumber \\ &&
\pm\frac{\kappa}{2}
\int d^{3}{\bf x}\int d^{3}\tilde{{\bf x}}
\Psi^{*}({\bf x};t)\Psi^{*}(\tilde{{\bf x}};t)
\delta^{(3)}({\bf x} -\tilde{{\bf x}})
\Psi({\bf x};t)\Psi(\tilde{{\bf x}};t),
\end{eqnarray}
which certainly satisfies the complex compatibility
conditions (\ref{compatibilitytwo}).

However, such equations do not respect the scaling
invariance $\Psi(t)\mapsto\lambda\Psi(t)$ that is
typical of the present theory. As Haag and Bannier (1978),
Kibble (1978), and Weinberg (1989) have suggested, this
scale invariance is physically desirable in order to
include separated systems properly. As later
emphasized by Jones (1994a), this same property
is responsible for the general possibility
of {\em nonlinear operators\/}, and thus a
nonlinear spectral theory --- {\em i.e. quantized
behaviour in a nonlinear theory\/}.

Demanding scale invariance at the level of the action, via
\begin{equation}
I[\lambda\Psi,\Psi^{*}]
= \lambda I[\Psi,\Psi^{*}]
=I[\Psi,\lambda\Psi^{*}],
\end{equation}
leads directly to (\ref{homogeneity}), once we
differentiate against $\lambda$. Thus
the Kibble--Weinberg formalism is fully
characterized as the {\em Complex
and Projective Hamiltonian Dynamics\/}.

That a mathematical structure of such importance
could go unrecognized for so long is surprising.
However, this is surely due to the subtleties of
complex geometry. The present formalism first
generalizes the classical fields of one space
coordinate into configuration space. This
entails a change in the conceptual stand taken,
but leaves the mathematics unaltered. This step
complete, we proceed to restrict the Lagrangian
dynamics to ensure compatibility with complex
numbers. At this stage one might be excused
for thinking a restricted mathematics would
be less frutiful. However, the demand of a
complex projective structure introduces
new structures that are not supported
by its general embedding.

The important conclusion for physics is that our
demand for a fully inclusive Lagrangian dynamics
for complex fields has here met with a  unique
solution! Thus we have identified the natural
mathematical system an inclusive nonlinear quantum
theory {\em must\/} employ if it is to recover
past successes. It is a remarkable thing to
enter upon a wider physical framework via
the mathematical restriction of the now
discarded classical theory.

The interesting novelty of this restriction
to complex fields is the natural occurrence
of nonlinear operators, via the presence of
non--bilinear hermitian forms.
To construct these we consider (\ref{homogeneity}),
as applied to the Hamiltonian functional, and so
obtain the canonical operator decomposition
\begin{equation}
\label{operator}
H[\Psi,\Psi^{*}] =
\int\int d^{3N}{\bf X}d^{3N}\tilde{{\bf X}}\,
\Psi^{*}(\tilde{{\bf X}};t)
\frac{\delta^{2} H[\Psi,\Psi^{*}]}
{\delta \Psi^{*}(\tilde{{\bf X}};t)
 \delta      \Psi({\bf X};t)}\Psi({\bf X};t),
\end{equation}
along with the subsidiary condition
\begin{equation}
\frac{\delta H[\Psi,\Psi^{*}]}
{\delta \Psi^{*}({\bf X};t)} =
\int d^{3N}\tilde{\bf X}\,
\frac{\delta^{2} H[\Psi,\Psi^{*}]}
{\delta \Psi^{*}({\bf X};t)\delta \Psi(\tilde{\bf X};t)}
\Psi(\tilde{\bf X};t).
\end{equation}
Thus homogeneous Hamiltonian functionals can always be
recast in the form of generalized expectation values,
where the matrix elements of the operator depend
upon $\Psi$. However, at each $\Psi$ these can
be diagonalized using the linear spectral
theory (Kreyszig 1989). Thus a complete
set of states is available to erect a
tangent Hilbert space at each point on
the manifold of all normalizable $\Psi$.

Combining these observations with the dynamical
equations (\ref{Wham}), one can now relate the
generalized framework with the familiar linear
dynamics, as traditionally expressed in Dirac
notation. Making the identifications
\begin{eqnarray}
\langle{\bf X},t|\Psi\rangle  &\leftrightarrow & \Psi({\bf X};t) \\
\langle{\bf X},t|\hat{H}[\Psi,\Psi^{*}]|\tilde{\bf X},t\rangle
&\leftrightarrow &
\frac{\delta^{2} H[\Psi,\Psi^{*}]}
{\delta \Psi^{*}({\bf X};t)\delta \Psi(\tilde{\bf X};t)} \\
\langle{\bf X},t|\hat{H}[\Psi,\Psi^{*}]|\Psi\rangle
&\leftrightarrow &
\frac{\delta H[\Psi,\Psi^{*}]}
{\delta \Psi^{*}({\bf X};t)},
\end{eqnarray}
the equation (\ref{Wham}) becomes (Kibble 1978)
\begin{equation}
\label{Kibble}
i\hbar\frac{d}{dt}|\Psi\rangle
 = \hat{H}[\Psi,\Psi^{*}]|\Psi\rangle,\;\;\mbox{and c.c.}
\end{equation}
which is the generalized Schr\"{o}dinger equation in operator form.

This is most interesting in connection with the spectral
theory of nonlinear dynamical systems. The eigenstates
of this formalism may be viewed as stationary points
of the flow in complex projective space. Then $\Psi$
is an eigenstate of its associated tangent--space
operator $\hat{H}[\Psi,\Psi^{*}]$. Physically,
we expect the time--dependence to read
\begin{equation}
\label{ansatz}
\Psi_{E}({\bf X};t) \equiv \Psi({\bf X};0)
\exp\left\{-\frac{i}{\hbar}Et\right\},
\end{equation}
in an eigenstate. Adopting the Rayleigh--Ritz variational
principle (Morse and Feschbach 1953), we look for critical
points of the normalized hamiltonian, i.e. we set
\begin{equation}
\label{eigenstate}
\frac{\delta }{\delta \Psi^{*}({\bf X};t)}
\left(\frac{H[\Psi,\Psi^{*}]}
{{\cal N}[\Psi,\Psi^{*}]}\right) = 0,
\end{equation}
where
$${\cal N}[\Psi,\Psi^{*}]
= \int d^{3N}{\bf X} \Psi^{*}({\bf X};t)\Psi({\bf X};t),$$
is the {\em norm--functional\/}. This fixes the
stationarity condition (Weinberg 1989)
\begin{equation}
\label{stationary}
\frac{\delta H[\Psi,\Psi^{*}] }
     {\delta \psi^{*}({\bf X};t)}
=    \left(\frac{H}{{\cal N}}\right) \psi({\bf X};t),
\end{equation}
with $E = H/{\cal N}$, the total energy in the
stationary state. In the physical language of states
and operators this is the familiar condition (now
intrinsically self--consistent):
\begin{equation}
\hat{H}[\Psi,\Psi^{*}]|\Psi\rangle = E |\Psi\rangle,
\end{equation}
which is the basis of nonlinear spectral theory (where
(\ref{homogeneity}) allows us to set ${\cal N}=1$). In
this manner one may verify (\ref{ansatz}) as being an
appropriate ansatz for nonlinear eigenfunctions.

\section{Empirical constraints upon nonlinearity}
The recent experiments of Bollinger et al. (1989) yielded an
upper bound of $4\times 10^{-27}$ for the relative magnitude
of nonlinear self--energy effects in freely precessing
Beryllium nuclei. This, and other similar null results
(for a review see Bollinger et al. (1992)), show
that single--particle nonlinearities are physically
uninteresting to contemplate.

However, these null results do not exclude the emergent
model of Penrose (1993), where quantum nonlinearity is
to intrude whenever a macrosopic
body, containing many particles, is split by an
amplified chain of interaction with a single
micro--particle into a gross superposed state.
Then the degree of nonlinearity felt by one
particle will be subtly dependent upon the
context into which it is placed, i.e. emergent effects are possible.

Therefore, we confine attention to the study of
many--body nonlinearities, and further look to
physical sources of self--interaction. Among
the four known physical interactions only
gravitation is not directly tested at the
quantum level. Although it is weak, it
does not screen, and so could play a
decisive role in collective
many--body physics. For our concern,
the main question is how quantum
gravity modifies dispersion.

\section{Cosmic localization and the observer}
Consider a universe containing just one scalar massive
neutral particle. In the linear theory dispersion will
spread the wave--packet, ultimately without limit. An
observer is then invoked to find the particle here,
or there, from time to time (Heisenberg 1949).
Thereafter, the observer assigns a new $\Psi$
to represent his or her knowledge about the
particle state. The localization achieved
is thus determined by the accuracy of
the measuring device.

\begin{figure}[p]
\centerline{\epsfbox{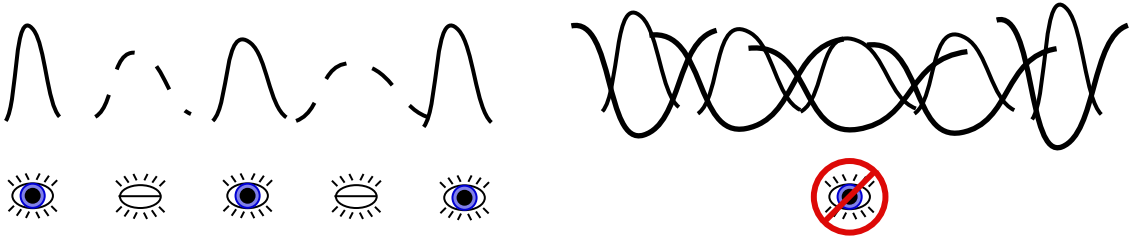}}
\caption{\label{cosmo}On a cosmic scale the Copenhagen theory is open. A
simple example is the Newtonian cosmology fixed by a scalar
massive field, with no other sources present. Some external
agent, here represented by an eyeball, must intervene to
define what happens through the act of observation, panel
(a). In the Schr\"{o}dinger theory the act of observation
is to be replaced by a dynamic localizing self--potential,
panel (b). In that case the cosmology is closed, and one
may legitimately pursue an objective interpretation.}
\end{figure}

Unfortunately, there is no measuring device out there
to observe the universe. Thus a Copenhagen physicist
must conceive of a {\em cosmic observer\/}, to sit
above all, and conjure from among all possibilites
the histories that are, see Fig.~\ref{cosmo}(a).
 Necessarily, one has passed
outside of physics at this point, which we
prefer to avoid. In place of the observer, in
place of the arbitrary selection of a cosmic
initial condition, one may give preference
to a theory which was explicit about what
an observer is (Bell 1973). Ideally, we
should make no presumption of sentience
(Wigner 1962), and so adhere to the
established model of matter ruled
by a few fundamental interactions,
independent of consciousness.

Within nonlinear theories the scope for treating
interactions is much wider. For instance, via a
potential $V[\Psi,\Psi^{*}]$ the wavefunction of
the universe experiences a separable dynamical
back--reaction. Such potentials may function
as a non--sentient observer to achieve the
necessary localization, see
Fig.~\ref{cosmo}(b). In this model all effects
now attributed to the observer should be
traced to nonlinear self--interactions.

Hence we seek to replace the Copenhagen obsever
by a localizing self--interaction. The options are: 1) strong force,
2) weak force, 3) electromagnetism, and
4) gravitation. Among these only gravitation
is not directly tested at the quantum level.
It is {\em universally attractive\/}, and
{\em additive\/}. Gravitational
localization grows stronger with
mass--density, whereas dispersion
decreases with mass. In the competition between effects
that dominate at either end of this spectrum one may
set a scale for the onset of emergent behaviour.

\section{Gravitational self--energy}
For the problem of one particle in an otherwise
source--free universe the only interaction that
is possible, and which may be plausibly invoked
to achieve both of the aforementioned aims, is
the gravitational self--energy due to the
stress--energy of its own wavefunction.

Ordinarily second--quantization is invoked to treat the physics
of self--energy. However, this approach has encountered severe
and persistent difficulties for gravitation. The field theory
is non--renormalizable and thus unpredictive (Isham 1992).
As noted previously by Barut (1990), the physical effect of
self--interaction can be modelled within nonlinear theories
using a self--potential $V[\Psi,\Psi^{*}]$. Then one need
not ``second--quantize'' gauge fields in order to equip
their sources with self--interaction. In place of the
traditional procedure we look for a self--potential
consistent with the continuum interpretation, and
corresponding with the classical treatment of
self--energy. The obvious choice is to take
\begin{equation}
\rho({\bf x};t) = m\psi({\bf x};t)\psi^{*}({\bf x};t)
\end{equation}
as the mass density for our particle. Then the physical
correspondence principle is met by adopting the Poisson
equation
\begin{equation}
\nabla^{2}\Phi_{\rm gravity}({\bf x}) = 4\pi G m\rho({\bf x};t),
\end{equation}
as its source.
Solving this we obtain the {\em gravitational self--potential\/}
\begin{equation}
\label{selfpot}
\Phi_{\rm gravity}({\bf x};t) = - G m\int
\frac{\psi^{*}(\tilde{\bf x};t)\psi(\tilde{\bf x};t)}
{|{\bf x} - \tilde{\bf x}|}\,d^{3}\tilde{\bf x},
\end{equation}
with the coupling strength fixed again by the correspondence
principle. Coupling (\ref{selfpot}) back upon the particle,
we compute its {\em gravitational self--energy\/}
\begin{equation}
\label{self}
E_{\rm gravity}
= - \frac{Gm^{2}}{2}
\int\int
\frac{\psi^{*}(\tilde{\bf x};t)\psi^{*}({\bf x};t)
          \psi(\tilde{\bf x};t)    \psi({\bf x};t)}
{|{\bf x} - \tilde{\bf x}|}\,d^{3}\tilde{\bf x} d^{3}{\bf x},
\end{equation}
where the factor $1/2$ avoids double counting. The result is
a non--perturbative and finite mass renormalization
$m\mapsto m + \delta m$, given by
$\delta m = E_{\rm gravity}/c^{2}$.

This choice recovers classical continuum results in a direct
manner. However, it remains to incorporate this term into a
wave--equation which recovers the Copenhagen free--particle
results when $E_{\rm gravity}$ may be regarded as negligble
relative to the wavepacket kinetic energy. This demand is
met once we choose the hamiltonian functional
\begin{eqnarray}
\lefteqn{H[\Psi,\Psi^{*}] =
\int \frac{\hbar^{2}}{2m}
\nabla\psi^{*}({\bf x};t)\cdot\nabla\psi({\bf x};t)\,d^{3}{\bf x}}
 \nonumber \\  & &
 - \frac{Gm^{2}}{2{\cal N}[\Psi,\Psi^{*}]} \int\int
\frac{\psi^{*}(\tilde{\bf x};t)\psi^{*}({\bf x};t)
          \psi(\tilde{\bf x};t)    \psi({\bf x};t)}
{|{\bf x} - \tilde{\bf x}|}
\,d^{3}\tilde{\bf x} d^{3}{\bf x},
\label{oneparticleenergy}
\end{eqnarray}
where scaling by $1/{\cal N}[\Psi,\Psi^{*}]$ in the second
term ensures that (\ref{homogeneity}) is satisfied. Using
(\ref{Wham}), we deduce the corresponding {\em gravitational
Schr\"{o}dinger equation}
\begin{equation}
\label{oneparticle}
i\hbar\frac{\partial}{\partial t}
\psi({\bf x};t) =
\left\{ -\frac{\hbar^{2}}{2m}\nabla^{2}
- - - \frac{G m^{2}}{{\cal N}}
\int \frac{\psi^{*}(\tilde{\bf x};t)\psi(\tilde{\bf x};t)}
{|{\bf x} - \tilde{\bf x}|}\,d^{3}\tilde{\bf x}
- - - \frac{E_{\rm gravity}}{\cal N}
\right\}\psi({\bf x};t).
\end{equation}
This generalizes the free--particle wave--equation
to include gravitational self--energy while ensuring that
we can recover the familiar Dirac formalism of linear
operators upon a Hilbert space via (\ref{operator}).
Also, the term $E_{\rm gravity}/{\cal N}$ ensures
that the Copenhagen rule for forming expectation
values will recover the self--energy functional
(\ref{oneparticleenergy}).

The main feature of interest is the existence of a spectral
theory for (\ref{oneparticle}) which runs analogous to that
for the hydrogen atom. Every wavefunction fixes
a self--potential, and the time--independent
Schr\"{o}dinger equation has a complete family of
eigenstates. Certain wavefunctions will then be
eigenstates of their own potential, and these
are stationary states of the system as a whole.
Thus we may reinterpret quantization as a
{\em nonlinear eigenvalue problem\/}
(cf. Schr\"{o}dinger 1928), and trace the
apparently discrete properties of nature
to the existence of quantized stationary states.

\section{Newtonian quantum gravity}
To obtain a consistent many--particle theory, in which
gravitation remains a non--entangled observer, we look
for a non--entangled treatment of mutual interaction.
It must meet the correspondence principle, and allow
for a consistent treatment of quantum statistics.

Here we take inspiration from the Hartree--Fock approximation
(Brown 1972), which provides a very simple physical model for
nonlinear mutual interactions. The method derives from a
variational principle (Kerman and Koonin 1976), and is
fully compatible with quantum statistics. It replaces
the usual two--body pairwise entangling interactions
by a sum of one--body non--entangling potentials. In
electromagnetism the source term for these is just
the one--body charge density. For electromagnetism
it is known to be approximate\footnote{For instance, the
entangling nature of electromagnetism is easily established
via spectral studies of many--electron atoms. In the earliest
calculations by Hartree (1928) he obtained energy levels that
differed from the experimental data by a few percent for the
lowest lying states. Indeed, Lieb and Simon
(1974) have since established that H--F energies
are generally larger than their linear Coulomb
counterparts, although for high--lying states
the predictions are asymptotically equal. Thus
entangled and non--entangled treatments of the
Coulomb interaction are distinguishable purely
via spectral studies.}, but
for gravitation there is no empirical data to check.

Guided thus, we postulate the many--body Hamiltonian
\begin{eqnarray}
\lefteqn{H^{(N)}[\Psi,\Psi^{*}] \equiv
\int d^{3N}{\bf X}\,\sum_{i=1}^{N}
\frac{\hbar^{2}}{2m_{i}}
\nabla_{{\bf x}_{i}}\Psi^{*}({\bf X};t)\cdot
\nabla_{{\bf x}_{i}}\Psi({\bf X};t) } \nonumber \\
& & - \frac{1}{{\cal N}[\Psi,\Psi^{*}]}
\int\int d^{3N}{\bf X}d^{3N}\tilde{\bf X}\,
\left(\frac{1}{2}\sum_{i,j=1}^{N}
\frac{Gm_{i}m_{j}}{|{\bf x}_{i} - \tilde{\bf x}_{j}|}\right)
\Psi^{*}({\bf X};t)
\Psi^{*}(\tilde{\bf X};t)
\Psi(\tilde{\bf X};t)
\Psi({\bf X};t),
\label{manybodyham}
\end{eqnarray}
which is compatible with quantum statistics. Applying
(\ref{Wham}) we obtain the equation of motion
\begin{equation}
\label{manybodydynamics}
i\hbar\frac{\partial}{\partial t}
\Psi({\bf X};t)  = \left\{-\sum_{i=1}^{N}
         \frac{\hbar^{2}}{2 m_{i}}\nabla^{2}_{{\bf x}_{i}}
        + \sum_{i=1}^{N} m_{i}\Phi({\bf x}_{i})
- - - \frac{E_{\rm gravity}}{{\cal N}[\Psi,\Psi^{*}]} \right\}
\Psi({\bf X};t).
\label{manyparticle}
\end{equation}
Here $\Phi({\bf x};t) $ is the gravitational potential
\begin{equation}
\label{Spotential}
\Phi({\bf x};t) =  - \sum_{i=1}^{N}
\frac{Gm_{i}}{{\cal N}[\Psi,\Psi^{*}]}
\int d^{3N}\tilde{\bf X}\,
 \frac{\Psi^{*}(\tilde{\bf X};t)\Psi(\tilde{\bf X};t)}
{|{\bf x} - \tilde{\bf x}_{i}|},
\end{equation}
which now depends upon only one--coordinate, and is
the same for each particle.

Since the one--body density (\ref{onebodydensity}) is
the source for the gravitational field this theory is
consistent with the intended physical interpretation.
Observable fields are one--body fields, and the
candidate observer monitors only these. Hence
we advance (\ref{manybodydynamics}) as a
plausible equation consistent with a
continuum quantum theory.

Obviously, a direct test of (\ref{manyparticle}) is out of
the question. Nevertheless, we can contrast it with the
canonically quantized theory specified by the Coulomb
potential
\begin{equation}
\label{copenhagen}
V({\bf X}) \equiv
- - -\frac{1}{2}\sum_{i\ne j;i,j=1}^{N}
\frac{Gm_{i}m_{j}}{|{\bf x}_{i} - {\bf x}_{j}|}.
\end{equation}
Just as with many--electron atoms, the gravitational
spectra of Copenhagen and Schr\"{o}dinger theories of
Newtonian gravity must differ (Lieb and Simon 1974).
Purely spectral studies of the bulk
excitations of a cold assembly of many neutral
particles, such as a dense Bose--Einstein
condensate, could arbitrate in favour of
either theory. Thus the {\em principles\/}
of Copenhagen physics are open to
{\em falsification\/} via tests
of the continuum alternative.

\begin{figure}[p]
\centerline{\epsfbox{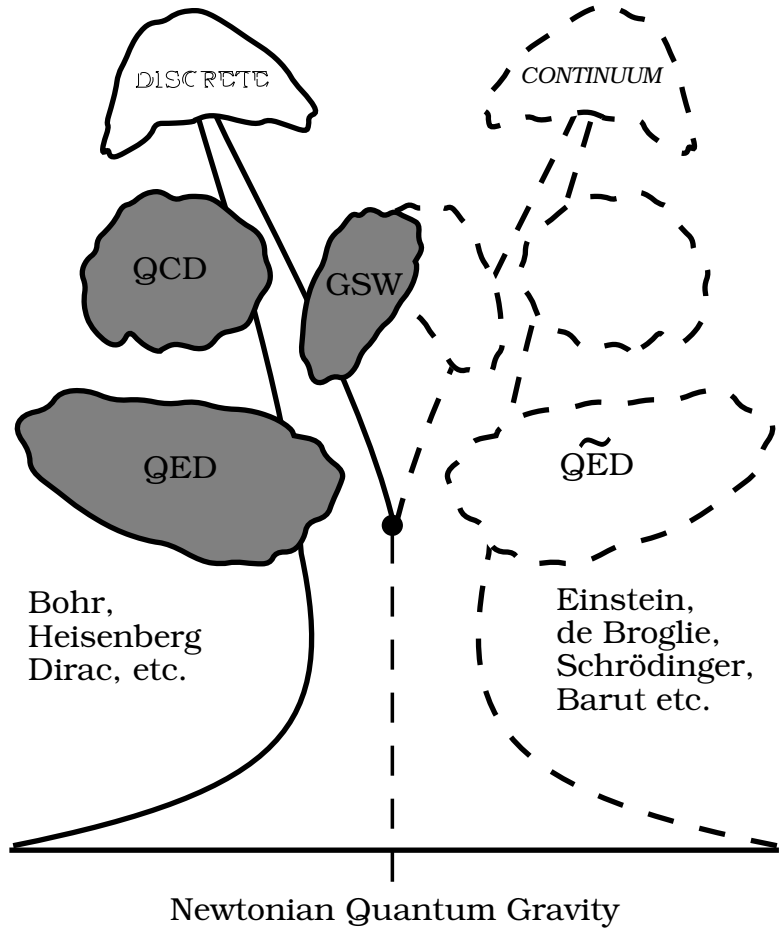}}
\caption{\label{forked}
Since there are two consistent ways to include gravitation,
and these differ in their philosophy,
physical interpretation, equations, and predictions, we may
view Newtonian quantum gravity as a critical fork which
divides quantum field theories into two classes. One is
a discrete theory, the other a continuum theory. Since
quantum gravity is not directly testable we must seek
indirect tests of the continuum alternative. The key
issue must be self--energy, and whether a consistent
nonlinear treatment is possible for the remaining
physical interactions. The self--field quantum
electrodynamics of Barut (1990), here labelled
as $\tilde{\rm QED}$, is an obvious candidate
theory for deeper studies in unification.}
\end{figure}

The significance of this simple observation cannot be
overstated. Since the treatment of self--energy
here adopted is consistent with that employed
by Barut (1990) in his nonlinear self--field
quantum electrodynamics, a self--consistent
non--perturbative continuum quantum field
theory is technically possible, see
Fig\ref{forked}. Furthermore, since he
found agreement with perturbative QED (at
least to $O(\alpha)$), one may take the
prospect of it {\em being correct\/}
rather seriously. Newtonian quantum
gravity thus represents a critical
fork in the development of quantum
field theories; it is the litmus
test of reality (Jones 1995).

\section{Localized solitary wave solutions}
As a first check upon the cosmical self--consistency of pure Newtonian
gravitation we search for bound and stable solitary wave solutions.
These are to set the scale for wave--packets in just the manner
that we presently conceive an external observer may do.

For simplicity, we will concentrate upon the case of $N$
identical bosons. The scenario of key interest is a
Bose--Einstein condensate described by the trial wavefunction
\begin{equation}
\label{trialwave}
\Psi({\bf X};t) =
\prod_{j=1}^{N} f({\bf x}_{j})\exp\{-iEt/\hbar\},
\end{equation}
consisting of $N$ bosons, each of mass $m$, with
$N$ identical wavefunctions $f({\bf x})$.
According to the prescription (\ref{Spotential})
the mass of this B--E condensate generates the
binding self--potential
\begin{equation}
\label{trialpot}
\Phi({\bf x}) =  - NGm\int d^{3}\tilde{\bf x}\,
 \frac{f^{2}(\tilde{\bf x})}
{|{\bf x} - \tilde{\bf x}|}
\end{equation}
(where we have set ${\cal N}[\Psi,\Psi^{*}] =1$).

Substituting (\ref{trialwave}) and (\ref{trialpot}) into
equation (\ref{manybodydynamics}), and taking variations
according to (\ref{stationary}), the original $N$--body
eigenvalue problem separates into $N$ copies
of the $1$--body equation
\begin{equation}
\label{bosonstar}
\nabla^{2}f({\bf x}) +
\frac{2m}{\hbar^{2}}\left\{\epsilon - m\Phi({\bf x})
\right\} f({\bf x}) = 0,
\end{equation}
coupled solely via the potential (\ref{trialpot}). With
the given B--E ansatz, (\ref{trialwave}), it suffices
to solve just one of these equations and so determine
the one unknown function $f$.

No analytical solutions to (\ref{bosonstar}) are known, but
numerical solutions are readily obtained. Indeed these have
been examined extensively in the literature of {\em boson
stars\/}, beginning with Ruffini and Bonnazola (1969),
followed by Thirring (1983), Friedberg et al. (1987),
and Membrado et al. (1989). In the previous studies
(\ref{manybodydynamics}) was interpreted as the
non--relativistic Hartree--Fock approximation
to the Copenhagen theory of quantum gravity,
as defined by (\ref{copenhagen}). Here we
view it as fundamental.

To discuss numerical solutions we must first define
the nonlinear eigenvalue $\epsilon$. Referring back
to (\ref{manybodydynamics}) we have
$$\epsilon \equiv E + E_{\rm gravity}/N,$$
where $E$ is the true one--particle eigenenergy,
and $E_{\rm gravity}$, is the total self--energy.
Thus $\epsilon$ is an eigenparameter, and is not
the physical eigenvalue.

Further, although the physical boundary conditions
are that $\Phi({\bf x})\rightarrow 0$ as ${\bf x}
\rightarrow \infty$ one can always redefine the
value of $\epsilon$ as follows:
$$\epsilon - m\Phi({\bf x})
= [\tilde{\epsilon} - \lambda]
- - - [m\Phi({\bf x}) - \lambda],$$
with $\lambda$ arbitrary. To fix it for numerical
computations we introduce a constraint due to
Membrado et al. (1989). The virial theorem
demands that $2 E_{\rm kinetic} = - E_{\rm gravity}$,
so that $E_{\rm total} =
E_{\rm kinetic} + E_{\rm gravity} = - E_{\rm kinetic} = + \frac{1}{2}E_{\rm
gravity}$.
Since $N\epsilon = E_{\rm total} + E_{\rm gravity}$, where
$E_{\rm total} = NE$, with $E$ the one--particle energies
above, we obtain the relation
\begin{equation}
\label{epsilon}
N\epsilon  = \frac{3}{2}E_{\rm gravity} = - 3 E_{\rm kinetic},
\end{equation}
showing that $\epsilon = 3E$, which is minus three times the
kinetic energy of a one--particle state. Since these are
readily computed independently of the numerical boundary
conditions imposed upon $\Phi$ the eigenvalues are then
fully determined.

Friedberg et al. (1987), have shown how the solutions
to (\ref{bosonstar}) form an  homologous family in
which all solutions are obtained as rescalings of certain
universal functions. Dividing (\ref{bosonstar}) through
by $f$ and taking the Laplacian of both sides we obtain
\begin{equation}
\label{reduced}
\nabla^{2}\left(\frac{\nabla^{2} f}{f}\right)
= \frac{8\pi}{a_{\rm g}}f^{2},
\end{equation}
and so pick out the {\em gravitational Bohr radius}
\begin{equation}
a^{(N)}_{\rm g} = \frac{\hbar^{2}}{G Nm^{3}},
\end{equation}
as the relevant scale parameter.

To solve for the spherically symmetric states ($S$--wave)
we introduce a radial coordinate $\rho$, and introduce
two universal functions $f^{\star}(\rho)$ and $g^{\star}(\rho) \equiv
\Phi^{*}(\rho) - \epsilon^{*}$ defined as solutions of the
system $\nabla^{2} f^{\star} = g^{\star}f^{\star}$ and
$\nabla^{2} g^{\star} = (f^{\star})^{2}$. Assuming we
have these one may check that the rescaled functions
\begin{eqnarray}
\label{funiversal}
f(r) & = & \frac{2^{1/2}}{\pi^{1/2}(\gamma_{1})^{2}(a^{(N)}_{\rm g})^{3/2}}
f^{\star}\left(\frac{2 r}{\gamma_{1}a^{(N)}_{\rm g}}\right)\\
\label{phiuniversal}
\Phi(r) & = & \frac{2}{(\gamma_{1})^{2}}
\frac{G^{2}Nm^{4}}{\hbar^{2}}
\left\{g^{\star}\left(\frac{2 r}
{\gamma_{1}a^{(N)}_{\rm g}}\right) + \epsilon^{*}\right\}
\end{eqnarray}
solve the system $\nabla^{2} f = (2m/\hbar^{2})gf$ and $\nabla^{2} g
= 4\pi GN m^{2} f^{2}$, (where $g=m\Phi$). Adopting the standardized
boundary conditions of Friedberg et al. (1987), namely
$f^{\star}(\rho) = 1$ with $\left.df^{\star}/d\rho\right|_{\rho = 0} = 0$,
$g^{\star}(0) = \gamma_{0}$, and $\left.dg^{\star}/d\rho\right|_{\rho = 0} =
0$,
one simply adjusts $\gamma_{0}$ by shooting to meet the demand
$\lim_{\rho\rightarrow\infty}f^{\star}(\rho)=0$. Once an
eigenfunction is found we may fix the parameters
\begin{eqnarray}
\gamma_{1} & \equiv &
\int_{0}^{\infty} [f^{\star}(\rho)]^{2}\rho^{2}\,d\rho,\\
\epsilon^{*}(n) & \equiv &
\frac{3}{\gamma_{1}}
\int_{0}^{\infty} [f^{\star}(\rho)]^{2}g^{\star}(\rho)\rho^{2}\,d\rho,
\end{eqnarray}
via numerical integration.

\begin{figure}[p]
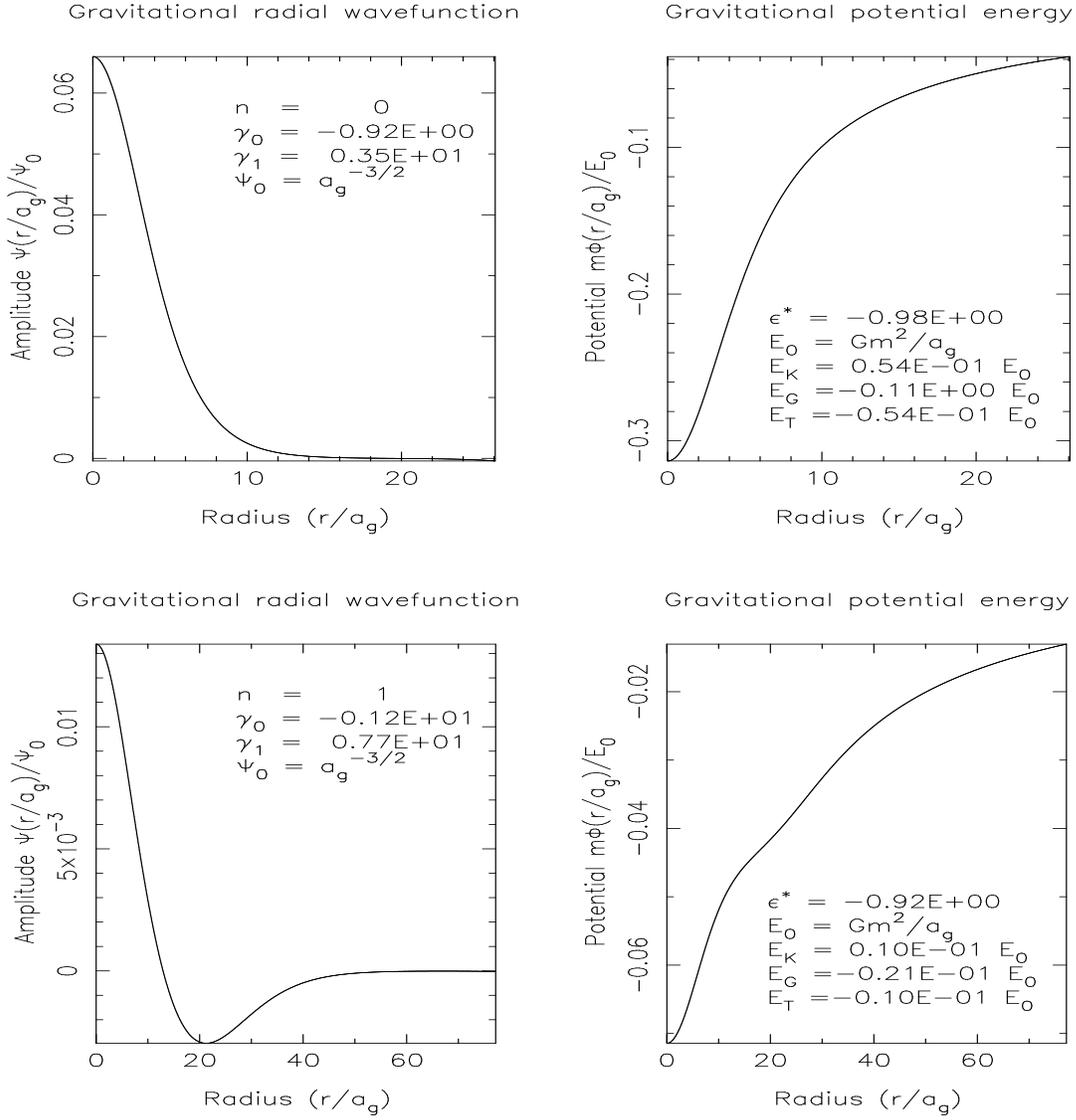

\centerline{\epsfbox{grav0.epsf}}
\centerline{\epsfbox{grav1.epsf}}
\caption{\label{figsolutions}Universal functions for (a) the gravitational
ground
state, and (b) ground--state self--potential; and (c) the first
excited--state, and (d) excited--state self--potential. A cosmic
scale is set for each wave--packet, and no external observer
is needed. Thus the gravitational \protect{Schr\"{o}dinger}
equation exhibits a closed Newtonian quantum cosmology.}
\end{figure}

The behaviour of solutions closely parallels that which
obtains with  linear eigenvalue problems such as
the hydrogen atom. Physical solutions occur labelled by
discrete values of $\gamma_{0}(n)$, with the principal
quantum number $n = 0,1,2,\ldots$ assigned by node
counting. Example solutions for the ground, and
excited state are shown in Fig.~\ref{figsolutions}.

The characteristic scale of the ground--state wave--packet
is around $10$ Bohr radii. This is an extremely interesting
number. For nucleon masses it is around $10^{23}$ m, but it
depends upon the number of particles present. If we take
$N=10^{23}$, i.e. around the Avagadro number, then the
cosmic scale of gravitational localization becomes $1$
m. This fact and the existence of excited states of
greater size indicates that, in continuum theories,
elementary ``particles'' are {\em dynamic entities\/}
whose size depends critically upon the experimental
context into which they are placed. This must be
expected in any departure from the Copenhagen
ideal of point--like objects since ``rigid''
models of particle structure are not
compatible with special relativity.

Indeed, other localizing, but entangling interactions, such as
electromagnetism, will alter the self--consistent solution that
determines these wavefunctions. As such, these computations
have no practical predictive content. They merely illustrate
how self--gravitation may equip nonlinear quantum theories
with a generic mechanism to set a fundamental scale for
wave--packet localization. Obviously, any predictions
for the onset of emergent behaviour must include the
effects of electromagnetism. In the static
approximation this is easy, we just include the
electromagenetic analogue of (\ref{copenhagen}).
However, the consistent treatment of radiative
processes is not straightforward. Indeed, some
way around the present canonical quantization
of the electromagnetic field would seem
essential. Again, the self--field QED of
Barut (1990) is an obvious candidate.

\section{Conclusion}
Nowadays it is often assumed that the principles of
Copenhagen physics are complete in every detail,
and that all new physics will conform to them.
Nevertheless, the problem of quantum gravity,
and the question of quantum measurement seem
always to elude a physical solution within
the orthodox scheme of thought.

Some years ago, Weinberg (1989) suggested that we
look to nonlinear theories as a means to test the
principle of linear superposition, the most central
tenet of Copenhagen quantum mechanics (Dirac 1958).
The stumbling block has been to locate new physical
principles compatible with nonlinearity. Only then
can we construct genuine alternative theories,
equal in their predictive power and manifest
self--consistency, but which finesse the
superposition principle via some subtle
and tellingly different predictions.

Here we have addressed the problem of interpreting
nonlinear theories in a predictive scenario that
is motivated by the outstanding difficulties in
quantum gravity, and quantum measurement. Thus
we approach the construction of a nonlinear
foil with clear physical problems in mind
- - --- we grant that the nonlinear option
may be {\em correct\/}.

The resulting theory differs in its underlying
principles in a clear manner: particles are to
be replaced by continuum fields; wavefunctions
are interpreted objectively using the device
of coarse--graining; observers are replaced
by a localizing self--interaction; jumps
are replaced by a two--step measurement
scenario; and the problem of quantized
values is re--interpeted as a nonlinear
eigenvalue problem.

The search for new principles represents a
logical game directed at locating a nonlinear
theory which has at least the same level of
scope and consistency as the Copenhagen
theory. This process led to the equation (\ref{manybodydynamics}),
as a consistent and predictive candidate. This equation
meets the two most precious demands of speculative physics:
it is falsifiable; and it is suggestive. Now that
we have a candidate theory, it is necessary to
develop these ideas further in the hope of
locating indirect experimental tests. Since
we have adopted a different treatment of
self--energy, it seems most natural to adopt
the unified treatment of electromagnetism
and gravitation as the critical question.
Here the work of Barut (1990) upon
an alternative non--perturbative version
of QED may serve as a useful model.

In summary, via study of the cosmical difficulties
posed by quantum measurement, and an appreciation
of the wider opportunities offered by nonlinear
theories, one may pass to a candidate theory
of nonlinear quantum gravity that is predictive,
supports a self--consistent interpretation, and
meets the demand of classical correspondence.
Most importantly,
the cosmological properties of the Copenhagen
and Schr\"{o}dinger theories differ greatly.
One is open, the other closed with respect
to the observer. This fact must determine
the eventual fate of an entire class of
relativistic theories. One has perhaps
a binary choice --- and some engaging
new questions to ask of nature.

\section{Acknowledgments}
I thank the Australian Research Council for their
support; Susan Scott and David McClelland, for a
very enjoyable meeting; Mark Thompson for some
helpful insights on B--E condensates; and many
other colleagues for useful discussions.

\section*{ References}
\noindent Schr\"{o}dinger, E. (1926). {\em Naturwiss.\/} {\bf 28}, 664; transl.
reprinted in
in Schr\"{o}dinger (1928), pp. 41--44.

\noindent Schr\"{o}dinger, E. (1928). `Collected Papers on Wave Mechanics'
(Blackie and Son).

\noindent Schr\"{o}dinger, E. (1935). {\em Naturwiss.\/} {\bf 23} 807; 823; and
844; transl. reprinted {\em In} `Quantum Theory and Measurement', (Eds W.H.
Zurek and J.A. Wheeler) pp. 152--167 (Princeton Univ. Press 1983).

\noindent Barut, A.O. (1988). {\em Ann. d. Phys.\/} {\bf 45}, 31.

\noindent Barut, A.O. (1990). {\em In\/} `New Frontiers in Quantum
Electrodynamics
and Quantum Optics', (Ed A.O. Barut) pp.345--370 (Plenum Press: New York).

\noindent Bell, J.S. (1973). {\em In} `The Physicist's Conception of Nature'
(Ed J. Mehra)
(Reidel: Dordrecht); reprinted in  Bell (1988), pp.40--44.

\noindent Bell, J.S. (1976). {\em In\/} `Quantum Mechanics, Determinism,
Causality and
Particles' (Eds M. Flato et al.) (D. Reidel: Dordrecht); reprinted in
Bell (1988), pp.93--99.

\noindent Bell, J.S. (1981). {\em In\/} `Quantum Gravity II', (Eds C.J.
Isham, R. Penrose and D. Sciama) (Clarendon Press: Oxford);
reprinted in  Bell (1988), pp.115--137.

\noindent Bell, J.S. (1988). `Speakable and Unspeakable in Quantum Mechanics',
(Cambridge Univ. Press).

\noindent Bohm, D. and Bub, J. (1966). {\em Rev. Mod. Phys.\/} {\bf 38}, 453.

\noindent Bohr, N. (1928). {\em Nature} {\bf 121},  580.

\noindent Bohr,  N. (1949). {\em In\/} `Albert Einstein, Philosopher ---
Scientist'
(Ed P.A. Schlipp) (Tudor: New York).

\noindent Bollinger, J.J.,  Heinzen, D.J., Itano, W.M.,  Gilbert, S.L., and
Wineland, D.J., (1989). {\em Phys. Rev. Lett.\/} {\bf 63}, 1031.

\noindent Bollinger, J.J.,  Heinzen, D.J., Itano, W.M.,  Gilbert, S.L., and
Wineland, D.J., (1992). {\em In\/} `Foundations of Quantum Mechanics'
(Proc. Sante Fe Workshop, Santa Fe, New Mexico, May 27--31, 1991)
(Eds T.D. Black et al.), p.40--46 (World Scientific: Singapore).

\noindent Brown, G.E., (1972). `Many--Body Problems' (North--Holland:
Amsterdam).

\noindent De Broglie, L. (1960). `Nonlinear Wavemechanics --- A Causal
Interpretation' (Elsevier).

\noindent Di\'{o}si, L. (1989). {\em Phys. Rev.\/} {\bf A40}, 1165.

\noindent Dirac, P.A.M. (1958). `The Principles of Quantum Mechanics', 4th edn.
(Oxford Univ. Press).

\noindent Einstein, A. (1956). `The Meaning of Relativity', 5th edn. p.165
(Princeton Univ. Press).

\noindent Friedberg, R., Lee  T.D., and Pang, Y. (1987). {\em Phys. Rev.\/}
{\bf D35}, 3640.

\noindent Gell--Mann, M. and Hartle, J.B. (1993). {\em Phys. Rev.\/} {\bf D47},
3345.

\noindent Haag, R. and Bannier, U. (1978). {\em Commun. Math. Phys.\/} {\bf
64}, 73.

\noindent Hartree, D.R. (1928). {\em Proc. Camb. Phil. Soc.\/} {\bf 24}, 89.

\noindent Heisenberg, W. (1949). `The Physical Principles of the Quantum
Theory', Chap. 4
(Dover Reprint: New York).

\noindent Isham, C.J. (1992). {\em In} `Recent Aspects of Quantum Fields',
(Springer Lecture Notes in Physics Vol. 396) (Eds H. Mitter and H. Gausterer),
pp. 123--225  (Springer--Verlag: Berlin).

\noindent Itzykson, C., and Zuber, J.--B. (1985). `Quantum Field Theory'
(McGraw--Hill).

\noindent Jones, K.R.W. (1992). {\em Phys. Rev.\/} {\bf D45}, R2590.

\noindent Jones, K.R.W. (1994a). {\em Ann. Phys. (N.Y.)\/}, {\bf 233}, 295.

\noindent Jones, K.R.W. (1994b). {\em Phys. Rev.\/} {\bf A50}, 1062.

\noindent Jones, K.R.W. (1995). {\em Mod. Phys. Lett.\/} {\bf A10}, 657.

\noindent Jordan, T.F. (1991). {\em Am. J. Phys.\/} {\bf 59}, 606.

\noindent Kerman, A., and Koonin, S.E. (1976). {\em Ann. Phys. (N.Y.)\/} {\bf
100}, 332.

\noindent Kibble, T.W.B. (1978). {\em Commun. Math. Phys.\/} {\bf 64}, 73.

\noindent Kibble, T.W.B. (1979). {\em Commun. Math. Phys.\/} {\bf 65}, 189.

\noindent Kreyszig, E. (1989). `Introductory Functional Analysis with
Applications',
pp.192--193 (Wiley: New York).

\noindent Lieb, E.H. and  Simon, B. (1974). {\em J. Chem. Phys.\/} {\bf 61}
735.

\noindent Membrado, M., Pacheco, A.F., and Sa\~{n}udo, Y. (1989).
{\em Phys. Rev.\/} {\bf A39}, 4207.

\noindent Morse, P.M., and Feshbach, H. (1953). `Methods of Theoretical
Physics: Part II', pp.1108--1119 (McGraw--Hill).

\noindent Penrose, R. (1993). {\em In\/} `General Relativity and Gravitation
1992',
(Proc. 13th Int. Conf. on General Relativity and Gravitation, Cordoba,
Argentina, June 28 --- July 4 1992) (Eds R.J. Gleiser, C.N. Kozameh
and O.M. Moreschi), pp. 179--189 (IOP: Bristol).

\noindent Ruffini R., and Bonazzola, S. (1969). {\em Phys. Rev.\/} {\bf 187},
1767.

\noindent Thirring, W. (1983). {\em Phys. Lett.} {\bf 127B}, 27.

\noindent Weinberg, S. (1989). {\em Ann. Phys. (N.Y.)\/} {\bf 194}, 336.

\noindent Wigner, E.P. (1962). {\em In\/} `The Scientist Speculates' (Ed R.
Good),
pp. 284--302 (Heinemann).

\noindent Zurek, W.H. (1981). {\em Phys. Rev.\/} {\bf D24}, 1516.

\noindent Zurek, W.H. (1982). {\em Phys. Rev.\/} {\bf D26}, 1862.

\noindent Zurek, W.H. (1991). {\em Phys. Today\/} {\bf 44}(10), 36.

\clearpage
\begin{figure}[p]
\centerline{\epsfbox{scenario.epsf}}
\caption{\label{scenario}In a nonlinear theory one must explain transition
probabilities without postulating them, so that a theory of
individual events is the natural goal. The existing theory
of environmental decoherence provides a route to justify
the values these take, see panel (a). There the effect
of the environment is to decohere the reduced density
matrix into diagonal form in the Zurek pointer basis
(Zurek 1981, 1982, 1991). However, because linear interactions
entangle states there is no way to cross--couple
the diagonal entries. If we invoke the Penrose
scenario (1993), of an emergent gravitational
instability, one might hope to reduce the
remaining terms in stochastic fashion
(cf. Di\'{o}si 1989). This scenario
is advanced as a guide to locating
candidate nonlinear terms.}
\end{figure}
\clearpage
\clearpage
\clearpage

\end{document}